\documentclass[aps,prl,twocolumn,superscriptaddress]{revtex4}

\usepackage{graphicx}

\begin{document}


\title{Electrical pump-and-probe study of spin singlet-triplet relaxation \\
in a quantum dot}

\author{S. Sasaki}
\email[]{satoshi@nttbrl.jp}
\affiliation{NTT Basic Research Laboratories, NTT Corporation, Atsugi-shi, Kanagawa 243-0198, Japan}

\author{T. Fujisawa}
\affiliation{NTT Basic Research Laboratories, NTT Corporation, Atsugi-shi, Kanagawa 243-0198, Japan}
\affiliation{Tokyo Institute of Technology, 2-12-1 Ookayama, Meguro-ku, Tokyo 152-8551, Japan}

\author{T. Hayashi}
\affiliation{NTT Basic Research Laboratories, NTT Corporation, Atsugi-shi, Kanagawa 243-0198, Japan}

\author{Y. Hirayama}
\affiliation{NTT Basic Research Laboratories, NTT Corporation, Atsugi-shi, Kanagawa 243-0198, Japan}
\affiliation{SORST-JST,  4-1-8 Honmachi, Kawaguchi, Saitama 331-0012, Japan}

\date{\today}

\begin{abstract}
Spin relaxation from a triplet excited state to a singlet ground state 
in a semiconductor quantum dot is studied 
by employing an electrical pump-and-probe method.
Spin relaxation occurs via cotunneling when the tunneling rate is relatively large,
confirmed by a characteristic square dependence of the relaxation rate
on the tunneling rate.
When cotunneling is suppressed by reducing the tunneling rate, 
the intrinsic spin relaxation is dominated by spin-orbit interaction.
We discuss a selection rule of the spin-orbit interaction 
based on the observed double-exponential decay of the triplet state.
\end{abstract}


\maketitle

\newcommand{\gt}{\Gamma_{\rm tot}}
\newcommand{\nt}{n_{\rm t}}
\newcommand{\tcot}{\tau_{\rm cot}}
\newcommand{\tso}{\tau_{\rm so}}
\newcommand{\twt}{t_{\rm h}}
\newcommand{\ts}{\tau_{\rm s}}

\begin{figure}
\includegraphics{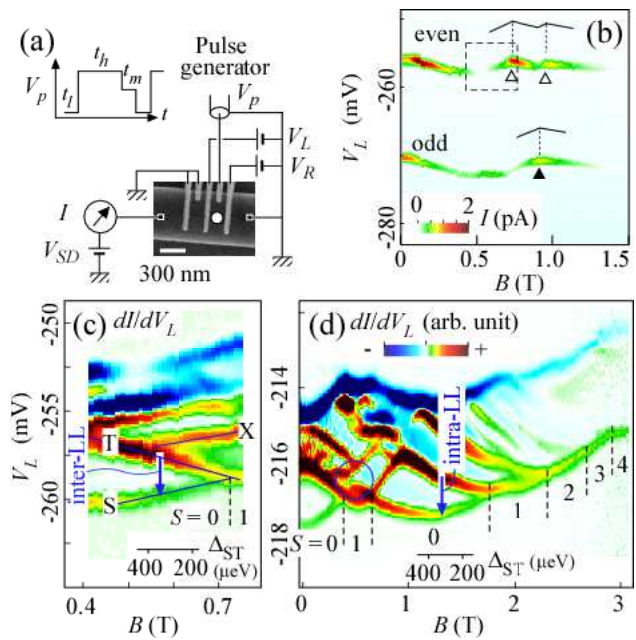}%
\caption{(a) SEM picture of the device together with
a schematic of the measurement set-up. 
(b) DC current, $I$, as a function of $V_{\rm L}$
and $B$ measured at $V_{\rm sd} = 0.15$~mV in the first cool-down.
$I$ for the even $N$ data is divided by 10.
Evolution of the current peak is sketched around the transition fields (triangles).
(c) $dI/dV_{\rm L}$ at $V_{\rm sd} = 1.2$~mV
taken at the dashed square region shown in (b).
$\Delta_{\rm ST}$ scale shows the region where the spin relaxation measurement is
conducted.
(d) $dI/dV_{\rm L}$ at $V_{\rm sd} = 1.0$~mV measured in the second cool-down.
The vertical dashed lines denote spin transitions of the ground state. 
A blue circle marks an anti-crossing between
two singlet states. Intensity for the high $B$ data is enhanced
by multiplying some smooth numerical function.
\label{f1}}
\end{figure}

Electron spin in semiconductors has been a focus of research 
 in the context of spintronics, in which 
spin is manipulated with spin-orbit coupling \cite{aws02, datta},
and of quantum computation, in which spin carries a quantum information \cite{loss98}.
In contrast to two-dimensional electron gas (2DEG) with continuum
density of states, electron spin in a quantum dot (QD) is basically free from
elastic scattering, and the resulting long-lived spin states are favorable
for spin-based applications.
Indeed, relaxation times of more than 100~$\mu$s have been reported in QDs
between Zeeman sublevels \cite{hanson03, elz04, krou}, 
as well as between a spin triplet and a singlet state \cite{fuji02, hanson04}. 
These relaxation processes have been discussed in terms of either
spin-orbit interaction or the cotunneling effect.
In this Letter, we study spin relaxation from a triplet state to 
a singlet state in a lateral QD, in which all the relevant parameters can be
controlled with the gate voltages.
We observe smooth transition of the relaxation mechanism
from the cotunneling regime to the spin-orbit regime by changing tunneling rates.
The decay of the excited triplet state follows a single exponential curve
in most conditions, but double exponential behavior is observed at a
particular magnetic field where the triplet state crosses another state.
This might be related to the long-lived spin-entanglement state
under strong spin-orbit interaction.

Figure \ref{f1}(a) shows a scanning electron micrograph (SEM)
image of our QD device. 
The \mbox{AlGaAs}/GaAs 2DEG is constricted 
by combined dry-etching and surface Shottky gates. 
We use only the three gates on the right-hand side  to form a single QD
as shown by the white circle. 
All the measurements are performed in a dilution refrigerator at $\sim$90~mK
with magnetic field, $B$, applied perpendicularly to the 2DEG. 

The dot used in this study has charging energy of $U \sim 2$~meV 
and electron number $N \sim 8$.
When the magnetic field is not very small ($B > 0.4$~T), 
electron orbitals in a QD can be classified by Landau level (LL) index, 
to which they approach in the high field limit \cite{nrc}. 
Many-body correction of direct and exchange Coulomb interactions induces 
spin and orbital transitions associated with different LLs in low magnetic field, 
but with the same LL in high magnetic field \cite{taru00}. 
Figure 1(b) shows an observed DC current, $I$, through the dot 
as a function of the left gate voltage, $V_{\rm L}$, and $B$. 
The two stripes show a pair-wise motion with $B$ reflecting spin degeneracy. 
The lower stripe, corresponding to odd $N$, involves 
a level crossing (denoted by a solid triangle), associated with two orbitals 
in different LLs. 
The upper stripe for even $N$ involves two spin transitions (open triangles) 
under Coulomb interactions. The ground state for even $N$ is assigned to be 
spin triplet between the two transition fields, 
otherwise spin singlet state \cite{taru00}. 
These spin states can be observed in the excitation spectrum of Fig.~\ref{f1}(c), 
in which the derivative of the current, $dI/dV_{\rm L}$, 
with a large $V_{\rm sd} = 1.2$~mV, is plotted as a function of $V_{\rm L}$ and $B$. 
Some excited states as well as the ground state that fall within the source-drain transport window are observed. 
We study spin relaxation from the triplet excited state to the singlet ground state 
(denoted by the arrow) separated by energy, $\Delta_{\rm ST}$.
This relaxation involves orbital change between different LLs (inter LL transition). 
Figure 1(d) is another excitation spectrum taken with the same device 
but in the second cool-down. 
In addition to the similar singlet-triplet transitions at $B \sim 0.5$~T, 
four spin-flip transitions are resolved at $B = 2 - 3$~T 
until the system enters a stable totally spin-polarized regime ($\nu = 1$), 
from which $N$ = 8 is estimated \cite{nrc}. 
We also study spin triplet-singlet relaxation (denoted by the arrow) 
that involves orbital change within the same LL (intra LL transition). 
The measurement in the 1st (2nd) cool-down with relatively fast (slow) cooling speed 
resulted in moderate tunnel rates and spin transition fields 
suitable for studying spin relaxation involving inter (intra) LL transition, 
but both sets of data show similar characteristics.

\begin{figure}
\includegraphics{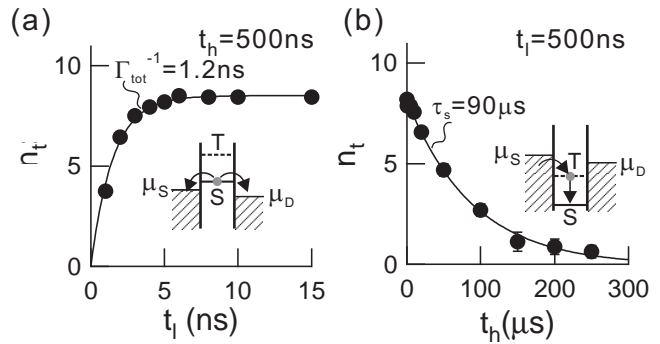}%
\caption{(a) $\nt$ as a function of the initialization time, $t_{\rm l}$, 
with $\Delta_{\rm ST}$ = 300~$\mu$eV ($B = 0.6$~T in the first cool-down).
The inset shows the energy diagram of the initialization process.
(b) $\nt$ as a function of the wait time, $\twt$.
The inset shows the energy diagram of the relaxation process.
\label{f2}}
\end{figure}

An electrical pump-and-probe measurement is performed by applying two-step square pulses 
to the plunger gate \cite{fuji02}. 
First, the singlet and triplet states are emptied by lifting both states
above the Fermi energy as shown in the inset to Fig. \ref{f2}(a) (initialization). 
The duration of this initialization is $t_{\rm l}$. 
Next, both states are pulled down below the Fermi energy 
as shown in the inset to Fig. \ref{f2}(b).
Then, only one electron can enter the dot because of the Coulomb blockade.
This electron, if it populates the triplet state with a probability, $P$,
is allowed to relax to the singlet ground state while the pulse height is
kept at this condition during the wait time, $\twt$. 
The three triplet sublevels with $S_{\rm Z} = \pm 1,\; 0$
are presumably populated with an equal probability
since the Zeeman splitting is negligibly small and the tunneling probability
does not depend on $S_{\rm Z}$.
Finally, the pulse height is adjusted so that only the triplet state is
within the transport window of 150 $\mu$eV defined by the Fermi energy of the
left and right leads (read-out). Then, the electron can contribute to the current
only if it remains in the triplet state after $\twt$. 
This read-out pulse width, $t_{\rm m}$, is fixed to 500~ns.
Actually, several electrons (1/(1-$P$)) flow during this time for the unrelaxed case.
Therefore, the average number of tunneling electrons per one pulse cycle,
$\nt$, follows an exponential decay $\nt={P \over 1-P} \exp(-\twt/\ts)$,
from which the spin relaxation time, $\ts$, can be determined \cite{fuji02}.

Figure 2(b) shows observed $\nt$ as a function of $\twt$ 
at $\Delta_{\rm ST}$ = 300~$\mu$eV. 
$\nt$ shows a single exponential decay with $\ts = 90$~$\mu$s. 
The relatively large $\nt (\twt =0) \simeq 8$, corresponding to $P \simeq 0.89$,
 comes from the fact that an injection into the triplet (singlet) state
is more (less) effective because an electron is added to the outer (inner) orbital 
with a larger (smaller) tunneling rate in this magnetic field region \cite{nrc}.

We can also determine the total tunneling rate, 
 $\gt$ (= $\Gamma_{\rm L}+\Gamma_{\rm R}$)
by changing the initialization pulse width, $t_{\rm l}$.
Here, $\Gamma_{\rm L}$ ($\Gamma_{\rm R}$) is the tunneling rate 
for the left (right) barrier, which is changed by $V_{\rm L}$ ($V_{\rm R}$).
Figure \ref{f2}(a) shows an example of such measurement. 
The rise time corresponds to the escape time from the singlet
ground state through both tunneling barriers, and 
$\gt$ is estimated by fitting the data to the expression
$\nt = {P \over 1-P} \{ 1-\exp (- t_{\rm l} \gt) \}$.
$\gt$ is successfully changed between 1$\times 10^8$ and 3$\times 10^9$~s$^{-1}$
by changing the gate voltages.

\begin{figure}
\includegraphics{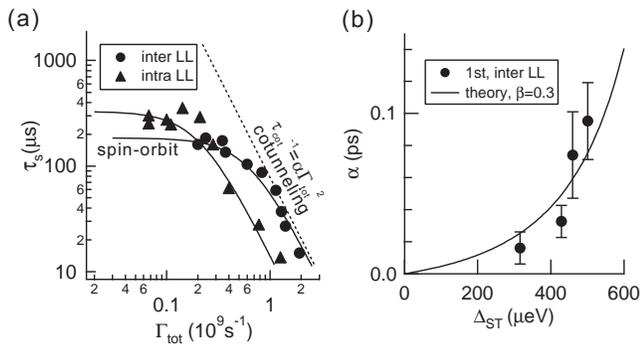}%
\caption{(a) Log-log plot of $\ts$ as a function of $\gt$. 
$\ts$ for inter LL is measured at $B = 0.6$ T ($\Delta_{\rm ST}$ = 300~$\mu$eV)
in the 1st cool-down, while that for intra LL is measured at $B = 1.2$ T 
($\Delta_{\rm ST}$ = 380~$\mu$eV) in the 2nd cool-down. 
The solid lines are fitted to the data.
(b) The coefficient $\alpha$ for the cotunneling component as a function of $\Delta_{\rm ST}$.
The curve is calculated with an effective tunneling rate, $\gt^*$ = $\beta \gt$.
\label{f3}}
\end{figure}

When $\gt$ is large,
higher-order tunneling, or cotunneling, is quite effective in causing an exchange of
electrons having opposite spins between the dot and the lead electrodes,
resulting in a spin relaxation. According to the second-order perturbation theory,
the cotunneling rate, $\tcot^{-1}$, is approximately given by
\begin{equation}
\tcot^{-1} = \Delta_{\rm ST} (\hbar \gt^\ast)^2 (\delta_-^{-1}+\delta_+^{-1})^2/h,
\label{eqcot}
\end{equation}
where $\delta_-$ and $\delta_+$ are energies required to excite the $N$ electron
triplet state to $N-1$ and $N+1$ electron virtual states, respectively \cite{fuji02}.
$\gt^\ast$ is the effective tunneling rate for the cotunneling process
from the triplet to the singlet state through either virtual state,
while the experimentally obtained $\gt$ measures the tunneling rate from the singlet
state to $N-1$ electron state.
Figure \ref{f3}(a) shows observed $\ts$ as a function of $\gt$ for representative
conditions. 
The data points at large $\gt$ are almost parallel with 
the dotted line having a slope $-2$,
{\it i.e.} \mbox{$\tcot ^{-1}= \alpha \gt^2$}, which is consistent with
 Eq.~(\ref{eqcot}), assuming a linear relation $\gt^\ast = \beta \gt$. 
Figure \ref{f3}(b) compares the observed $\Delta_{\rm ST}$ dependence of 
$\alpha$ with calculated 
$\alpha =  \{h \Delta_{\rm ST} (\delta_-^{-1}+\delta_+^{-1})^2 \beta^2\} / 4 \pi^2$.
We take $\delta_- = \delta_+ = (U/2 - \Delta_{\rm ST})$ in the calculation,
which approximates the experimental conditions. 
$\beta$ of 0.3 gives a reasonable fit to the experimental results.
Therefore, spin relaxation in the large $\gt$ regime can be well explained
by the standard cotunneling theory.

It is seen in Fig.~\ref{f3}(a) that, when $\gt$ is reduced, $\ts$ increases and 
eventually saturates. In this regime, the cotunneling process is suppressed, and
inelastic spin relaxation is dominated by phonon emission under the spin-orbit coupling effect rather than by coupling to nuclear spins, etc \cite{khae00,hanson04,krou}.
The solid lines in Fig.~\ref{f3}(a) are the curves (1/$\tso$+1/$\tcot$)$^{-1}$ fitted to the data.
Here, $\tso$ is the relaxation time due to the spin-orbit interaction, which is independent of $\gt$.

\begin{figure}
\includegraphics{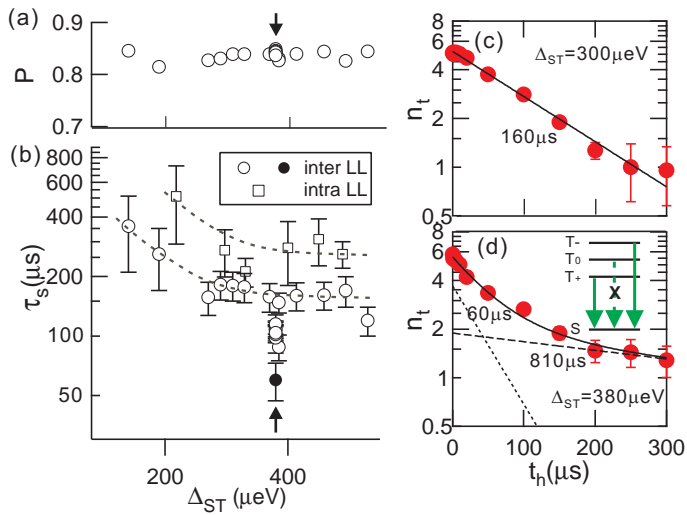}%
\caption{(a) $\Delta_{\rm ST}$ dependence of the excitation probability, $P$, 
for the inter LL relaxation (1st cool-down).
(b) $\Delta_{\rm ST}$ dependence of the $\ts$ 
for the inter and intra LL relaxations.
Data with open symbols are obtained by fitting a single exponential function,
while the solid circle represents the fast component of the double exponential function.
The dotted lines are guides for the eye.
(c) A logarithmic plot of $\nt$ vs. $\twt$ with 
$\gt = 2.0 \times 10^8$~s$^{-1}$ at $\Delta_{\rm ST}$ = 300~$\mu$eV ($B = 0.6$~T).
The solid line is an exponential function fitted to the data.
(d) A logarithmic plot of $\nt$ vs. $\twt$ with 
$\gt = 2.7 \times 10^8$~s$^{-1}$ at $\Delta_{\rm ST}$ = 380~$\mu$eV ($B = 0.55$~T). 
The solid line is a double-exponential function fitted to the data.
The dotted and dashed lines are the fast and slow components, respectively.
The inset schematically shows allowed and forbidden transitions from the
triplet sublevels to the singlet state.
\label{f4}}
\end{figure}

Figure \ref{f4}(b) shows $\Delta_{\rm ST}$ dependence of $\ts$ 
measured in the small $\gt$ regime where spin-orbit interaction is dominant. 
$\ts$ is almost constant in a wide
$\Delta_{\rm ST}$ regime (except at a dip around $\Delta_{\rm ST} \sim 380 \mu$eV)
and tends to increase when $\Delta_{\rm ST} < 200$~$\mu$eV 
for both inter and intra LL data.
This feature might have arisen from the phonon emission spectra in a QD
in the presence of spin-orbit coupling. The phonon emission rate
is maximized when the phonon
wavelength is comparable to the dot size (phonon energy of 300~$\mu $eV for the
dot size of 30~nm in the crystal growth direction) \cite{bock90, fuji02}. 
The similarity of the $\Delta_{\rm ST}$ dependence of the inter and intra LL data
supports this crude model.
Longer $\ts$ is observed for the intra LL case than for the inter LL case
in the whole $\Delta_{\rm ST}$ region explored. 
This might reflect the different orbital quantum numbers involved in each case, 
which are relevant to the orbital effect on the phonon emission and spin-orbit 
interaction \cite{dest04, bula}.
However, the precise mechanism is not clear yet. 

Generally, in a one-electron system, 
the spin-up (down) state of one orbital is coupled with the spin-down (up) state
of the other orbital when spin-orbit coupling is considered  
between the two orbitals \cite{bula}.
This coupling gives rise to finite phonon emission
probability between Zeeman sublevels.
In the case of two-electron singlet/triplet states, singlet state ($|S\rangle $)
is coupled with two of the triplet sublevels ($|T_+\rangle $ and $|T_-\rangle $)
having $S_{\rm Z} = \pm 1$ but not with the other sublevel ($|T_0\rangle $)
having $S_{\rm Z} = 0$ \cite{khae00, dick03}. 
Therefore, relaxation from the triplet to the singlet state
should have a selection rule in which $|T_0\rangle $ state is still free from the 
spin-orbit relaxation mechanism as shown in Fig.~\ref{f4}(d). 
This simple argument applies when the singlet-triplet energies are so close
to each other that coupling with other states is negligible. 
Unfortunately, our pump-and-probe technique is not available 
for the small $\Delta_{\rm ST}$ regime
because the minimum energy resolution is about 100~$\mu$eV.
When another singlet excited state is involved, however,
clear selectivity may appear in the vicinity of the level crossing
between the triplet and the singlet excited states.

Actually, as shown in Fig.~\ref{f1}(c),
we do see a level crossing with an unknown state (X) at about 
$B \simeq 0.52$~T ($\Delta_{\rm ST} \simeq 400$~$\mu$eV)
that could be a singlet state.
When the dot potential has no rotational symmetry, large anti-crossing is expected
between the {\it same} spin states \cite{anti}.
Typical anti-crossing energy between the same spin states in our
non-circular dot device is about 150~$\mu$eV 
(for instance, marked by the circle in Fig.~\ref{f1}(d)).
The unresolved anti-crossing between T and X states ($< 100 \mu$eV)
implies that X is a singlet state.
As shown by the arrow in Fig.~\ref{f4}(b), a sharp dip in $\ts$ is
observed at $\Delta_{\rm ST} \simeq 380 \mu$eV close to the X-T crossing point.
The width of the dip is as narrow as about 20~$\mu$eV in $\Delta_{\rm ST}$
($\sim 10$~mT in $B$) \cite{com2}.
The dip could be attributed to strong spin-orbit coupling around the crossing point,
resulting in the short relaxation time, as theoretically predicted in Ref.~\cite{bula}.
It should be noted that the decay of the pulse-induced current shows
a non-single exponential behavior around the dip as shown in Fig.~\ref{f4}(d),
while single exponential decay is always observed at other conditions, e.g., 
at $\Delta_{\rm ST} = 300 \mu$eV shown in Fig.~\ref{f4}(c).
The decay characteristic in Fig.~\ref{f4}(d) can be very well fitted 
with a double exponential funcion (the solid line);
$C_1 \exp (-\twt/\tso) + C_2 \exp (-\twt/\tcot)$. 
Here, $C_1$, $C_2$ and $\tso$ are fitting parameters, 
and $\tcot$ is an input parameter (= 810~$\mu$s) determined by
extrapolating the $\gt$ dependence of $\ts$ in the cotunneling regime 
to the present value of $\gt$. 
The fast component of the double exponential decay can be assigned to the
relaxation from $|T_+\rangle $ and $|T_-\rangle $ via spin-orbit coupling,
while the slow component to the relaxation from $|T_0\rangle $
via higher order spin-orbit coupling or the remaining cotunneling contribution.
We find that the obtained ratio of $C_1/C_2$ is 2.0 and $\tso$ is 60~$\mu$s. 
The ratio $C_1/C_2$ obtained at slightly different $\gt$ ranges between 1.6 and 2.0. 
These values of $C_1/C_2$ are close to 2, which is expected for 
an equal population of the three triplet sublevels. 

The above observations agree well with the selection rule 
for spin-orbit coupling that is enhanced in the vicinity of the X-T crossing. 
However, we cannot safely rule out other
possibilities like populating an X state in addition to the triplet state of interest.
Indeed, a very small increase ($\simeq 1 \%$) in the excitation probability $P$ is noted
at around $B = 0.55$~T as shown by the arrow in Fig.~\ref{f4}(a),
which could be due to injection into the X state. 
However, this effect is too small to explain the observed double exponential
behavior. Another possibility might be coupling to nuclear spins,
which often appears at the level coincidence of different spin states \cite{ono04}.

In summary, we have studied spin relaxation dynamics from the triplet excited state to 
the singlet ground state in a lateral QD. 
The dominant spin relaxation mechanism is 
cotunneling at a large tunneling rate, and it changes to
 spin-orbit interaction when cotunneling is suppressed.
The observed double exponential decay characteristic 
could reflect the selection rule for the singlet-triplet transition mediated by
spin-orbit interaction. Further investigation is required to prove this is the case.

We can think of an ``entanglement generator'' using this selection rule:
The singlet ground state $\left|S\right\rangle =\left| \uparrow \right\rangle _{a}\left| \downarrow
\right\rangle _{a}$ holds a spin pair in an orbital $a$, while the triplet state
contains non-entangled states $\left| T_{+}\right\rangle =\left| \uparrow
\right\rangle _{a}\left| \uparrow \right\rangle _{b}$ and $\left|
T_{-}\right\rangle =\left| \downarrow \right\rangle _{a}\left| \downarrow
\right\rangle _{b}$, and entangled state $\left| T_{0}\right\rangle =\frac{1%
}{\sqrt{2}}(\left| \uparrow \right\rangle _{a}\left| \downarrow
\right\rangle _{b}+\left| \downarrow \right\rangle _{a}\left| \uparrow
\right\rangle _{b})$, with an electron in each of the orbital $a$ and $b$. 
At a proper waiting time after the electron
injection [e.g. $\twt=$ 300~$\mu$s in the case of Fig.~\ref{f4}(d)], the system
is left in the entangled triplet state $|T_{0}\rangle $ with a probability
of 18 \% [$ \nt (\twt =300 \mu {\rm s}) \times (1-P)$] or otherwise in the
singlet ground state $|S\rangle $. Our pulse measurement is based on the
extraction of an electron from the unrelaxed $|T_{0}\rangle $ state
(the outer orbital with the high probability $P$), and
thus this scheme can be used to generate or analyze an entangled spin pair
by detecting the extracted electron with a sensitive
electrometer \cite{elz04, hanson04}. 

\begin{acknowledgments}
The authors acknowledge valuable discussions with Dr. A. V. Khaetskii and Y. Tokura. 
The work is financially supported by SORST-JST and Grant-in-Aid for
Scientific Research from the Japan Society for the Promotion of Science.

\end{acknowledgments}


\begin{thebibliography}{99}
\bibitem{com2}
Because of the narrow feature, the dip was discovered only in the 1st cool-down, 
but reproduced well during the same cool-down.
\end{thebibliography}


\end{document}